\documentclass[twocolumn,showpacs,amssymb,aps,prc,nofootinbib,floatfix,superscriptaddress]{revtex4-1}
\usepackage{amsmath,graphicx,color,ulem}
\usepackage{hyperref}
\newcommand{\trento}{T$\mathrel{\protect\raisebox{-2.1pt}{R}}$ENTo}

\begin{document}
\title{Understanding the correlation between elliptic and triangular flow}

\author{Mubarak Alqahtani}
\affiliation{Department of Physics, College of Science, Imam Abdulrahman Bin Faisal University, Dammam 31441, Saudi Arabia  }
\affiliation{Basic and Applied Scientific Research Center, Imam Abdulrahman Bin Faisal University, Dammam 31441 Saudi Arabia}
\author{Jean-Yves Ollitrault}
\affiliation{Universit\'e Paris Saclay, CNRS, CEA, Institut de physique th\'eorique, 91191 Gif-sur-Yvette, France}

\begin{abstract} 
 The relative correlation between the magnitudes of elliptic flow ($v_2$) and triangular flow ($v_3$) has been accurately measured in nucleus-nucleus collisions at the LHC collider. 
 As a function of the centrality of the collision, it changes sign and varies non-monotonically.
 We  show that this is naturally explained by two combined effects.
 The first effect is a skewness in initial-state fluctuations, which is quantified by the correlation between the geometry-driven elliptic deformation in the reaction plane and the fluctuation-driven triangularity $\varepsilon_3$.  
 We introduce an intensive measure of this skewness, which is generically of order unity and varies little above 5\% centrality.  
 We evaluate its magnitude using Monte Carlo simulations of the initial state, which show that it is sensitive to the nucleon width. 
 The second effect is the fluctuation of impact parameter relative to centrality classifiers used by experiment. 
 The ATLAS collaboration uses two different centrality classifiers, the multiplicity $N_{ch}$ and the transverse energy $E_T$. 
 We fit both sets of results for Pb+Pb collisions up to $\approx 40\%$ centrality with a single parameter, the intensive mixed skewness. 
 Its value inferred from experiment agrees with theoretical expectations.
\end{abstract}
\maketitle

\section{Introduction}
\label{s:introduction}

Elliptic flow, $v_2$~\cite{STAR:2000ekf,ALICE:2010suc}, and triangular flow, $v_3$~\cite{Alver:2010gr} are the hallmarks of the formation of a fluid~\cite{Schenke:2010rr} in ultrarelativistic nucleus-nucleus collisions. 
They are usually thought of as independent phenomena: 
Elliptic flow mostly originates from the almond geometry of the overlap area between colliding nuclei in non-central collisions~\cite{Ollitrault:1992bk}, while triangular flow is solely due to event-by-event fluctuations in the initial geometry~\cite{Alver:2010gr}. 
The naive expectation is that they are uncorrelated. 
But  the relative correlation between $v_2^2$ and $v_3^2$, defined by\footnote{
``{\rm nsc}'' stands for ``normalized symmetric cumulant''~\cite{ALICE:2016kpq}, the subscript $2,3$ refers to the Fourier harmonics involved ($v_2$ and $v_3$), and $\{4\}$ refers to the fact that the numerator is analyzed with a 4-particle cumulant~\cite{Borghini:2001vi}.}
\begin{equation}
  {\rm nsc}_{2,3}\{4\}\equiv \frac{\langle v_2^2 v_3^2\rangle-\langle v_2^2\rangle\langle v_3^2\rangle}{\langle v_2^2\rangle\langle v_3^2\rangle}
  \label{defnsc}
\end{equation}
(where angular brackets denote an average over events in a centrality class), has been measured in Pb+Pb collisions at the LHC~\cite{ALICE:2016kpq,ALICE:2021adw} and in Au+Au collisions at RHIC~\cite{STAR:2018fpo}, and is usually negative. 
This anti-correlation is also seen in hydrodynamic~\cite{Gardim:2016nrr,Zhu:2016puf,Schenke:2019ruo} and transport~\cite{Nasim:2016rfv,Li:2021nas} simulations, yet a simple picture of its origin is still missing. 

\begin{figure*}[th!]
    \includegraphics[width=\linewidth]{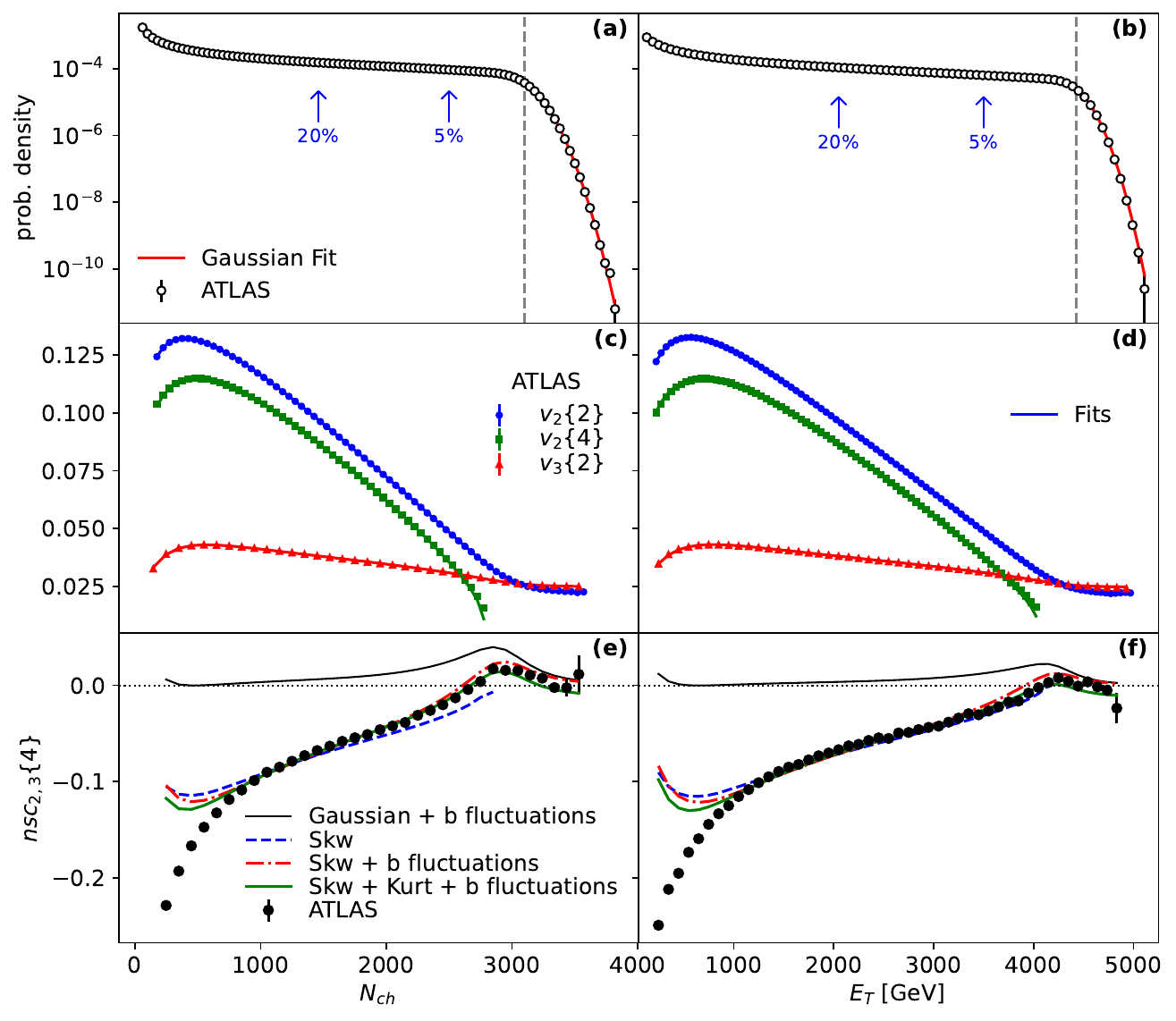}
    \caption{(a) and (b): Histograms of the two centrality classifiers used by ATLAS~\cite{ATLAS:2019peb}:  (a) Multiplicity $N_{ch}$ of charged particles with $p_T>0.5$~GeV/$c$ and $|\eta|<2.5$, where $p_T$ is the transverse momentum and $\eta$ the pseudorapidity; (b) Transverse energy $E_T$ in $3.2<|\eta|<4.5$. 
    Arrows indicate specific values of the centrality fraction for reference. 
      Solid lines are fits assuming Gaussian fluctuations at fixed impact parameter $b$~\cite{Das:2017ned}, and the vertical dashed line is the position of the knee, defined as the expectation value of the centrality classifier at $b=0$. The region around the knee corresponds to ultracentral collisions.  
      (c), (d): Symbols: $v_2\{2\}$, $v_2\{4\}$ and $v_3\{2\}$ data~\cite{ATLAS:2019peb}. Lines: fits taking into account impact parameter ($b$) fluctuations~\cite{Roubertie:2025qps}. 
      (e), (f): ATLAS data on ${\rm nsc}_{2,3}\{4\}$ data~\cite{ATLAS:2019peb}. 
      Dashed lines: One-parameter fits with skewness and without $b$ fluctuations (Eq.~(\ref{nscng})). 
      Thin solid lines: Calculated value (not a fit) without skewness and kurtosis, but taking into account $b$ fluctuations (Sec.~\ref{s:centrality}). 
      Dot-dashed lines: One-parameter fits with skewness {\it and\/} $b$ fluctuations. 
      Full lines: Two-parameter fits with skewness, kurtosis and $b$ fluctuations. 
}      
        \label{fig:atlasfits}
\end{figure*}

We focus on the detailed analysis of $ {\rm nsc}_{2,3}\{4\}$ carried out by the ATLAS Collaboration~\cite{ATLAS:2019peb} which uses a very fine centrality binning, and two different centrality classifiers, the multiplicity $N_{ch}$ around midrapidity, and the transverse energy $E_T$ deposited in forward and backward calorimeters (Fig.~\ref{fig:atlasfits} (e) and (f)). 
 At small $N_{ch}$ or  $E_T$, ${\rm nsc}_{2,3}\{4\}$ is negative. Then it increases towards central collisions where it becomes positive, reaches a maximum, and eventually decreases.
%JY: I removed "towards zero" because it actually goes to a small negative value, which is better seen as a function of ET, and which we understand as the effect of kurtosis. 
Our goal is to understand these data on the basis of general arguments, without invoking a specific hydrodynamic model. 

We first recall in Sec.~\ref{s:hydroresponse} the standard picture of hydrodynamic response relating $v_2$ and $v_3$ to the initial anisotropies $\varepsilon_2$ and $\varepsilon_3$. 
In Sec.~\ref{s:epsfluctuations}, we discuss event-by-event fluctuations, which we divide into classical and quantum fluctuations. 
Quantum fluctuations are local fluctuations of the initial density profile, which can be studied using standard tools of statistical physics. 
In particular, the moments of $\varepsilon_2$ and $\varepsilon_3$ can be systematically  decomposed into a hierarchy of cumulants~\cite{Kubo:1962dyl} whose magnitude follows general scaling laws. 
We show that ${\rm nsc}_{2,3}\{4\}$ is driven by non-Gaussian fluctuations, in the form of a mixed skewness and a mixed kurtosis.
In Sec.~\ref{s:trento}, we define intensive measures of these non-Gaussianities~\cite{Giacalone:2020lbm,Alqahtani:2024ejg,Roubertie:2025qps} which are expected to have reduced dependence on system size~\cite{ALICE:2023tej,ATLAS:2024jvf}. 
We evaluate their magnitudes using Monte Carlo simulations with the \trento{} model of initial conditions~\cite{Moreland:2014oya}. 
In Sec.~\ref{s:nongaussian}, we fit the intensive skewness to ATLAS data on ${\rm nsc}_{2,3}\{4\}$.
In Sec.~\ref{s:centrality}, we show that the finite centrality resolution~\cite{Das:2017ned} explains the positive ${\rm nsc}_{2,3}\{4\}$ in ultracentral collisions, as well as the differences between results based on $N_{ch}$ and $E_T$. 
Our results are summarized in Sec.~\ref{s:discussion}.

\section{Linear response to initial anisotropies}
\label{s:hydroresponse}

Throughout this paper, we assume that particle emission is described by the flow paradigm~\cite{Luzum:2011mm,Ollitrault:2023wjk}. 
Namely, particles in each event are emitted independently according to an underlying probability distribution. 
Let $P(\varphi)$ denote the azimuthal dependence of this probability distribution.
The complex anisotropic flow $V_n$ of the event is defined as its Fourier coefficient~\cite{Luzum:2013yya}:
\begin{equation}
  V_n\equiv \int_{0}^{2\pi}e^{in\varphi}P(\varphi)d\varphi. 
\label{defvn}
\end{equation}
In hydrodynamics, $V_n$ is to a good approximation proportional~\cite{Gardim:2011xv,Niemi:2012aj} to the complex anisotropy~\cite{Qiu:2011iv} 
$\varepsilon_n=\varepsilon_{n,x}+i\varepsilon_{n,y}$ for $n=2,3$~\cite{Teaney:2012ke} ($\varepsilon_n$ is the $n^{\rm th}$ Fourier coefficient of the transverse density profile, weighted with the $n^{\rm th}$ power of the distance $r$ to the center of energy): 
\begin{equation}
\label{hydroresponse}
V_n\approx\kappa_n\varepsilon_n,
\end{equation}
where $\kappa_n$ is a real linear response coefficient which is the same for all events in a centrality class. 
A global comparison between theory and ATLAS data gives the preferred values (Fig.~28 of Ref.~\cite{Nijs:2020roc})
\begin{align}
\kappa_2\approx 0.326\nonumber\\
\kappa_3\approx 0.235\label{trajectum}
\end{align}
for mid-central Pb+Pb collisions at $2.76$~TeV per nucleon pair. 
Throughout this paper, we assume that these relations hold, so that anisotropic flow and initial anisotropies are related by simple proportionality factors. 

If one injects Eq.~(\ref{hydroresponse}) into Eq.~(\ref{defnsc}), the response coefficient $\kappa_n$ cancels between the numerator and the denominator: 
\begin{equation}
  {\rm nsc}_{2,3}\{4\}= \frac{\langle\varepsilon_2 \varepsilon_2^*\varepsilon_3\varepsilon_3^*\rangle-\langle\varepsilon_2 \varepsilon_2^*  \rangle\langle\varepsilon_3\varepsilon_3^*  \rangle}{\langle \varepsilon_2\varepsilon_2^*\rangle\langle \varepsilon_3\varepsilon_3^* \rangle},
  \label{defnsceps}
\end{equation}
where $\varepsilon_n^*$ denotes the complex conjugate of $\varepsilon_n$. 

Therefore, a straightforward way of comparing ATLAS data with theory is to run a state-of-the-art Monte Carlo model of initial conditions, evaluate $\varepsilon_2$ and $\varepsilon_3$ in each event, and compute ${\rm nsc}_{2,3}\{4\}$ by averaging over events in a centrality bin.\footnote{A boost-invariant two-dimensional model suffices to study data as a function  of $N_{ch}$, but the $E_T$-based data require a proper modeling of the rapidity dependence, which is not always well reproduced~\cite{Yousefnia:2021cup}.}
Instead of this brute-force approach, we choose to carry out, in Secs.~\ref{s:epsfluctuations} and \ref{s:trento}, a general study of $\varepsilon_n$ fluctuations. 
We identify the leading terms contributing to ${\rm nsc}_{2,3}\{4\}$, which can be characterized with a few parameters. 

\section{Cumulants of initial anisotropies in the intrinsic frame}
\label{s:epsfluctuations}

Both the modulus and the phase of $\varepsilon_n$ fluctuate event by event. 
Event-by-event fluctuations fall broadly into two categories~\cite{Roubertie:2025qps}:
Classical fluctuations and quantum fluctuations. 
For a collision between spherical nuclei, the only parameter with negligible quantum uncertainty is the impact parameter.\footnote{We only study  spherical nuclei. For strongly deformed nuclei such as $^{238}U$, the orientation of the nucleus at the time of the collision is also a quasi-classical parameter, which must be fixed in order to properly define the intrinsic frame.}
%Add reference for deformed nuclei!!
Fluctuations of the magnitude $b$ and direction of impact parameter are classical in nature~\cite{Samanta:2023amp}. 
All the remaining fluctuations are quantum. 
They originate from the wavefunctions of colliding nuclei, or from the collision process itself. 

We first assume that $b$ is the same for all events in a centrality class or, equivalently, that the centrality determination is perfect.  
This assumption will be relaxed in Sec.~\ref{s:centrality}. 
We choose a coordinate system where the $x$ axis is the direction of impact parameter.\footnote{The $(x,z)$ plane is the reaction plane, and the $x$ axis is often called the reaction plane by a slight abuse of language.} 
We call this the ``intrinsic frame''~\cite{Alqahtani:2024ejg,Roubertie:2025qps}. 
It is easily accessible in numerical simulations, not in an actual experiment. 
Events in the intrinsic frame only differ by local density fluctuations, which originate from  quantum fluctuations.

Moments of $\varepsilon_2$ and $\varepsilon_3$ can be decomposed into cumulants which isolate the connected part of the correlation to all orders~\cite{Borghini:2000sa}. 
For the moments of order $2$  appearing in the denominator of Eq.~(\ref{defnsceps}), this decomposition takes the form: 
\begin{align}
\langle \varepsilon_2\varepsilon_2^*\rangle &=\langle \varepsilon_2\rangle\langle \varepsilon_2^*\rangle+\langle \varepsilon_2\varepsilon_2^*\rangle_c\nonumber\\
&=\bar\varepsilon_2^2+\sigma_{\varepsilon_2}^2,  \label{eps22}
\end{align}
where $\bar\varepsilon_2\equiv\langle\varepsilon_2\rangle$ is the mean elliptic deformation. 
$\bar\varepsilon_2$ is real because symmetry with respect to the $x$ axis amounts to the exchange $\varepsilon_n \leftrightarrow \varepsilon_n^*$. 
It vanishes for $b=0$ by azimuthal symmetry. 
The subscript $c$ in the last term of Eq.~(\ref{eps22}) denotes the cumulant, that is, the variance of $\varepsilon_2$, which we denote by $\sigma_{\varepsilon_2}^2$.\footnote{Moments and cumulants coincide to order $1$: $\langle\varepsilon_2\rangle_c=\langle\varepsilon_2\rangle$.}

The same decomposition applies to $\varepsilon_3$, with one simplification:
Near mid-rapidity, the symmetry under the exchange of target and projectile implies symmetry under $\varphi\to\varphi+\pi$, which changes $\varepsilon_3$ into $-\varepsilon_3$. 
Therefore, the average triangularity $\langle \varepsilon_3\rangle$ vanishes~\cite{Alver:2010gr} and 
\begin{equation}
\langle \varepsilon_3\varepsilon_3^*\rangle =\langle \varepsilon_3\varepsilon_3^*\rangle_c=\sigma_{\varepsilon_3}^2. \label{eps32} 
\end{equation}

We now decompose the moment of order four appearing in the numerator of Eq.~(\ref{defnsceps}) into all possible lower-order clusters, represented by cumulants. 
Cumulants which are odd in $\varepsilon_3$, such as $\langle \varepsilon_2^*\varepsilon_3\rangle_c$, vanish by symmetry, and the non-trivial terms are: 
\begin{align}
\langle \varepsilon_2\varepsilon_2^* \varepsilon_3\varepsilon_3^*\rangle =&
\langle \varepsilon_2\rangle\langle\varepsilon_2^*\rangle\langle \varepsilon_3\varepsilon_3^*\rangle_c+
\langle \varepsilon_2\varepsilon_2^*\rangle_c\langle \varepsilon_3\varepsilon_3^*\rangle_c\nonumber\\
&+2\langle \varepsilon_2\rangle\langle\varepsilon_2^* \varepsilon_3\varepsilon_3^*\rangle_c+
\langle \varepsilon_2\varepsilon_2^* \varepsilon_3\varepsilon_3^*\rangle_c.\label{4moment} 
\end{align}
The last line involves cumulants of order 3 and 4, respectively, where the order is the number of terms in the product. 

When fluctuations result from $N$ independent sources (here, an order of magnitude of $N$ is given by the number of participant nucleons~\cite{Miller:2007ri}), a cumulant of order $k$ is of order $N^{1-k}$~\cite{Borghini:2000sa}.
For $k=1$, the only non-vanishing cumulant is the mean anisotropy $\bar\varepsilon_2$ which depends on the shape, but is independent of $N$. 
The cumulants of order $k=2$ are the variances $\sigma_{\varepsilon_n}^2$ which are of order  $1/N$~\cite{Bhalerao:2006tp,PHOBOS:2007vdf}.

For systems with large $N$, such as Pb+Pb collisions, the magnitude of cumulants rapidly decreases as the order $k$ increases. 
The central limit theorem consists of neglecting cumulants beyond order $2$, in which case fluctuations of $\varepsilon_n$ are Gaussian~\cite{Voloshin:2007pc}. 
The last line in Eq.~(\ref{4moment}) vanishes, and the moment reduces to: 
\begin{equation}
\langle \varepsilon_2\varepsilon_2^* \varepsilon_3\varepsilon_3^*\rangle =
(\bar\varepsilon_2^2+\sigma_{\varepsilon_2}^2)\sigma_{\varepsilon_3}^2=\langle \varepsilon_2\varepsilon_2^*\rangle \langle \varepsilon_3\varepsilon_3^*\rangle 
\label{4momentgauss} 
\end{equation}
This implies that ${\rm nsc}_{2,3}\{4\}$ defined by Eq.~(\ref{defnsceps}) vanishes. 

Therefore, in order to explain the non-zero correlation seen experimentally, one must go beyond the central limit theorem and include cumulants beyond order $2$, which represent non-Gaussian fluctuations~\cite{Bhalerao:2019fzp}.\footnote{Non-Gaussian fluctuations of the fluctuations of a single harmonic $v_n$ are well documented~\cite{Yan:2013laa,Yan:2014afa}. 
They explain the non-zero $v_3\{4\}$ seen in Pb+Pb~\cite{ALICE:2011ab} and Xe+Xe~\cite{CMS:2019cyz,CMS:2025xtt} collisions , the hierarchy of higher-order cumulants of $v_2$ ($v_2\{4\}$, $v_2\{6\}$, $v_2\{8\}$) in p+Pb collisions~\cite{CMS:2015yux,CMS:2019wiy}, as well as the small splitting between $v_2\{4\}$ and $v_2\{6\}$ in mid-central Pb+Pb collisions~\cite{Giacalone:2016eyu,CMS:2017glf,ALICE:2018rtz,CMS:2023bvg}.}
Keeping all the terms in Eq.~(\ref{4moment}), one obtains the expression
\begin{equation}
  {\rm nsc}_{2,3}\{4\}= \frac{2\bar\varepsilon_2\langle\varepsilon_2^* \varepsilon_3\varepsilon_3^*\rangle_c+
\langle \varepsilon_2\varepsilon_2^* \varepsilon_3\varepsilon_3^*\rangle_c}
  {\langle \varepsilon_2\varepsilon_2^*\rangle\langle \varepsilon_3\varepsilon_3^* \rangle}.
  \label{nscnobfluct}
\end{equation}
The leading term  in the numerator is the first term, which involves a cumulant of order $3$. 
We refer to this cumulant as to the ``mixed skewness''. 
The second term is a mixed kurtosis~\cite{Bhalerao:2019fzp} which is suppressed by an additional power of $1/N$. 
It matters only for central collisions where the mixed skewness vanishes by azimuthal symmetry~\cite{PHOBOS:2007vdf}. 

The smaller the system, the larger the non-Gaussian corrections. 
The very fact that ${\rm nsc}_{2,3}\{4\}$ becomes larger, in absolute magnitude, for small $N_{ch}$ (Fig.~\ref{fig:atlasfits} (e)) hints at generic non-Gaussian corrections. 
This trend is confirmed by other collaborations, who report a large negative ${\rm nsc}_{2,3}\{4\}$  in low-multiplicity Pb+Pb collisions~\cite{ALICE:2019zfl} and in p+Pb collisions~\cite{CMS:2017kcs}. 
We now introduce intensive measures of the mixed skewness and kurtosis, which have reduced dependence on the system shape and size. 

\section{Intensive measures of the mixed skewness and kurtosis}
\label{s:trento}

\subsection{Definitions}

The mean anisotropy $\bar\varepsilon_2$ measures the shape of the initial density profile, while the magnitude of eccentricity fluctuations $\sigma_{\varepsilon_n}$  (with $n=2$ or $n=3$), which is of order $N^{-1/2}$, gives a measure of the size $N$. 

The mixed skewness $\langle\varepsilon_2^* \varepsilon_3\varepsilon_3^*\rangle_c$ is a Fourier coefficient of order $2$, which vanishes for central collisions, like the mean anisotropy $\bar\varepsilon_2$. 
One therefore expects that the ratio $\langle\varepsilon_2^* \varepsilon_3\varepsilon_3^*\rangle_c/\bar\varepsilon_2$ has reduced dependence on the shape parameter $\bar\varepsilon_2$. 
No such scaling is required for the mixed kurtosis $\langle \varepsilon_2\varepsilon_2^* \varepsilon_3\varepsilon_3^*\rangle_c$, which is azimuthally isotropic. 

We then suppress the dependence of $N$, which is $1/N^2$ for the skewness and $1/N^3$ for the kurtosis, by scaling them with appropriate powers of $\sigma_{\varepsilon_2}$ and $\sigma_{\varepsilon_3}$. 
We define the intensive mixed skewness $\Gamma_S$ and the intensive mixed kurtosis $\Gamma_K$ by: 
 \begin{align}
 \Gamma_S&\equiv\frac{\langle\varepsilon_2^* \varepsilon_3\varepsilon_3^*\rangle_c
 }{\bar\varepsilon_2\sigma_{\varepsilon_2}(\sigma_{\varepsilon_3})^3}\nonumber\\
  \Gamma_K&\equiv\frac{\langle \varepsilon_2\varepsilon_2^* \varepsilon_3\varepsilon_3^*\rangle_c}{(\sigma_{\varepsilon_2})^3(\sigma_{\varepsilon_3})^3}.
\label{v2v3cumul}
 \end{align} 
Note that there is some arbitrariness on how one distributes the factors of $\sigma_{\varepsilon_2}$ and $\sigma_{\varepsilon_3}$ in the denominator, since both are of similar magnitude. 
We have chosen a definition such that $\Gamma_K$ is symmetric under the exhange of $\varepsilon_2$ and $\varepsilon_3$. 

\begin{figure}[t]
    \includegraphics[width=\linewidth]{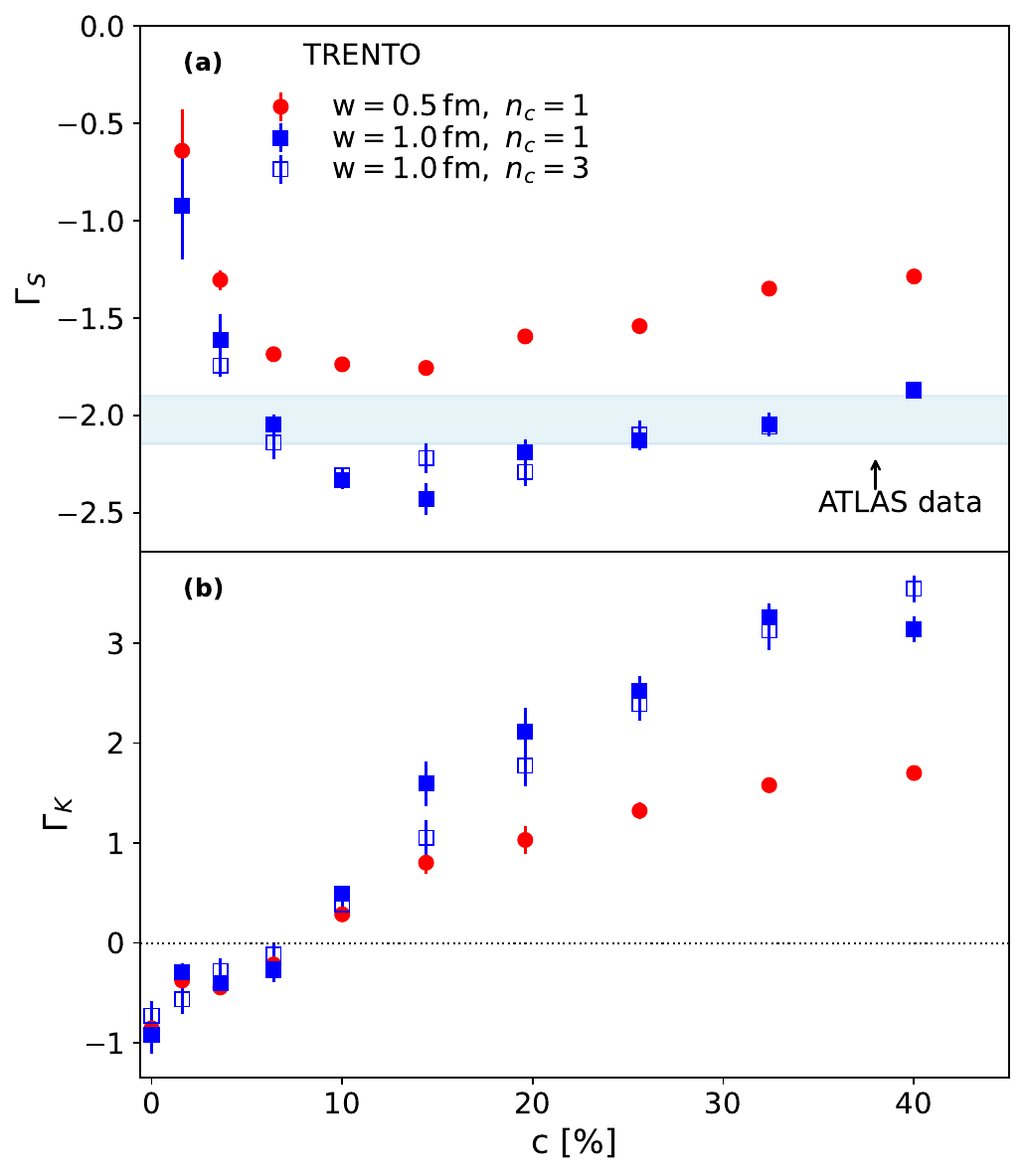}
    \caption{Intensive mixed skewness $ \Gamma_S$ (a) and mixed kurtosis $\Gamma_K$ (b), defined by Eqs.~(\ref{v2v3cumul}) and (\ref{defmixedskewness}), as a function of the centrality fraction $c$ in \trento{} simulations. 
    We define $c\equiv \pi b^2/\sigma_{\rm PbPb}$, where $\sigma_{\rm PbPb}\approx 785$~fm$^2$ is the total inelastic Pb+Pb cross section (see Sec.~\ref{s:centrality})
    The band in panel (a) is the range of values inferred by fitting ATLAS data on ${\rm nsc}_{2,3}\{4\}$ with impact parameter fluctuations taken into account (dot-dashed and full lines in  Fig.~\ref{fig:atlasfits} (e) and (f), see Sec.~\ref{s:centrality}). 
        \label{fig:trento}}
\end{figure}
%legend missing for the shaded band in figure

\subsection{Monte Carlo simulations}

In order to evaluate the signs and orders of magnitude of $\Gamma_S$ and $\Gamma_K$, and their centrality dependence, we use the \trento{} model~\cite{Moreland:2014oya}, which simulates the entropy density profile in the transverse plane $s(x,y)$ with event-by-event fluctuations. 
The model has several free parameters:  The nucleon width $w$, the number of constituents $n_c$ in each participant nucleon~\cite{Moreland:2018gsh}, their width $v$ (for which we choose the default value $v=0.5$~fm), and a fluctuation parameter $k$ which determines how the thickness of a given constituent fluctuates. 
Thickness functions $T_A(x,y)$ and $T_B(x,y)$ are obtained by summing the thicknesses of participants in each nucleus.
The entropy density is finally modeled as $s\propto\sqrt{T_AT_B}$, which results in a good global description of observables after the hydrodynamic evolution~\cite{Moreland:2018gsh,JETSCAPE:2020mzn,Nijs:2020roc}. 
We compare three different sets of parameters: 
We run the model without nucleon substructure with  $w=0.5$~fm and $w=1$~fm, and with $n_c=3$ constituent quarks and $w=1$~fm. 
For each set of parameters, we tune the fluctuation parameter $k$ in such a way that the relative standard deviation of $N_{ch}$ (which we assume to be proportional to the entropy) at $b=0$ matches the value inferred from ATLAS data~\cite{Yousefnia:2021cup}, namely, $4.5\%$.

We run the model for several fixed values of impact parameter.
For each event, we evaluate $\varepsilon_2$ and $\varepsilon_3$. 
The mixed skewness and kurtosis are then obtained by expressing the cumulants in terms of moments, which are evaluated by averaging over events: 
\begin{align}
\langle\varepsilon_2^* \varepsilon_3\varepsilon_3^*\rangle_c&=
\langle \varepsilon_{2,x}|\varepsilon_3|^2\rangle-\langle \varepsilon_{2,x}\rangle\langle|\varepsilon_3|^2\rangle\nonumber\\
\langle \varepsilon_2\varepsilon_2^* \varepsilon_3\varepsilon_3^*\rangle_c &=\langle |\varepsilon_2\varepsilon_3|^2\rangle-\langle|\varepsilon_2|^2\rangle\langle|\varepsilon_3|^2\rangle-2 \bar\varepsilon_{2}\langle\varepsilon_2^* \varepsilon_3\varepsilon_3^*\rangle_c. 
\label{defmixedskewness}
\end{align}
The first line shows that the mixed skewness is the linear correlation between the anisotropy in the reaction plane $\varepsilon_{2,x}$ (i.e., the real part of $\varepsilon_{2}$) and the squared modulus of the triangularity $|\varepsilon_3|^2$. 
We then evaluate $\Gamma_S$ and $\Gamma_K$ using Eq.~(\ref{v2v3cumul}), where $\sigma_{\varepsilon_2}$ and $\sigma_{\varepsilon_3}$ are obtained using Eqs.~(\ref{eps22}) and (\ref{eps32}). 

The resulting values of  $\Gamma_S$ and $\Gamma_K$ are plotted in Fig.~\ref{fig:trento} as a function of the centrality fraction $c$. 
Both are of order unity as expected. 
$\Gamma_S$ is negative, and its value depends somewhat on the nucleon width $w$, not on the number of constituent quarks.
It varies abruptly for small $c$. 
It is a feature which we do not understand, but we argue that it is of limited relevance. 
First, note that both the numerator and the denominator of $\Gamma_S$ vanish in the limit $c\to 0$, which is the reason why error bars grow. 
The important quantity is the numerator, which enters Eq.~(\ref{nscnobfluct}). 
For small $c$, the contribution of $\Gamma_S$ in Eq.~(\ref{nscnobfluct}) is suppressed by a factor $c^2$. 
As we shall see in Secs.~\ref{s:nongaussian} and \ref{s:centrality}, the effects of the skewness become larger as the centrality fraction grows. 
The important feature is that the centrality dependence of $\Gamma_S$ is modest in the centrality region where skewness matters, that is, excluding the most central collisions. 

On the other hand, $\Gamma_K$ changes sign from negative to positive as the impact parameter increases.\footnote{
Interestingly, the intensive skewness and kurtosis of the $\varepsilon_2$ distribution display similar features, illustrated in Fig.2 of Ref.~\cite{Roubertie:2025qps}.}
The negative sign for central collisions was already reported in Ref.~\cite{Bhalerao:2019fzp}. 
We will check in Sec.~\ref{s:centrality} that the effect of the kurtosis is much smaller than that of the skewness for most centralities. 
Unlike the skewness, the kurtosis mostly matters for ultracentral collisions, where it becomes the leading non-Gaussian correction. 

\section{Mixed skewness from ATLAS data}
\label{s:nongaussian}

We now investigate to what extent ATLAS data on  ${\rm nsc}_{2,3}\{4\}$ can be explained with a centrality-independent intensive skewness $\Gamma_S$. We neglect the kurtosis, which is a cumulant of higher order. 

We first need to evaluate $\bar\varepsilon_2$, $\sigma_{\varepsilon_2}$ and $\sigma_{\varepsilon_3}$ as a function of centrality. 
They are inferred from data on $v_2$ and $v_3$, using the response relations Eq.~(\ref{hydroresponse}). 
 $\sigma_{\varepsilon_3}$ is obtained using Eq.~(\ref{eps32}): 
\begin{equation}
\sigma_{\varepsilon_3}=\langle \varepsilon_3 \varepsilon_3^*\rangle^{1/2}=\frac{1}{\kappa_3} \langle v_3^2\rangle^{1/2}=\frac{1}{\kappa_3} v_3\{2\}, 
\label{v32}
\end{equation}
where the rms $v_3$, denoted by $v_3\{2\}$, is measured experimentally. 
Similarly, Eq.~(\ref{eps22}) gives: 
\begin{equation}
(\bar\varepsilon_2^2+\sigma_{\varepsilon_2}^2)^{1/2}=\frac{1}{\kappa_2} v_2\{2\}. 
\label{v22}
\end{equation}
Separating the mean anisotropy $\bar\varepsilon_2$ from the fluctuations $\sigma_{\varepsilon_2}$ is not as straightforward, since the reaction plane is not known in experiment. 
One uses the information on $v_2\{4\}$, which is defined by~\cite{Borghini:2001vi}: 
\begin{align}
v_2\{4\}&\equiv \left( 2\langle v_2^2\rangle^2-\langle v_2^4\rangle\right)^{1/4}\nonumber\\
&=\kappa_2 \left( 2\langle \varepsilon_2\varepsilon_2^*\rangle^2-\langle \varepsilon_2\varepsilon_2\varepsilon_2^*\varepsilon_2^*\rangle\right)^{1/4}
\label{v24}
\end{align}
Decomposing the moment of order $4$ in the right-hand side into lower-order cumulants and assuming Gaussian fluctuations, i.e., keeping only cumulants up to order 2, one obtains:\footnote{In the decomposition, we neglect the term $\langle \varepsilon_2\varepsilon_2\rangle_c\langle \varepsilon_2^*\varepsilon_2^*\rangle_c$.  This contribution is the square of the asymmetry of fluctuations (the difference between projections onto $x$ and $y$ axes), whose effect is known to be small~\cite{Roubertie:2025qps}.} 
\begin{align}
\langle \varepsilon_2\varepsilon_2\varepsilon_2^*\varepsilon_2^*\rangle
&=\langle\varepsilon_2\rangle^4+4\langle\varepsilon_2\rangle^2\langle \varepsilon_2\varepsilon_2^*\rangle_c+ 2\langle \varepsilon_2\varepsilon_2^*\rangle_c^2\nonumber\\
&=\bar\varepsilon_2^4+4\bar\varepsilon_2^2\sigma_{\varepsilon_2}^2+2\sigma_{\varepsilon_2}^4, 
\end{align}
Injecting into Eq.~(\ref{v24}) and using Eq.~(\ref{eps22}), one obtains the simple relation~\cite{Voloshin:2007pc}
\begin{equation}
\bar\varepsilon_2 =\frac{1}{\kappa_2}v_2\{4\}. 
\label{v24gauss}
\end{equation}

We then use Eq.~(\ref{nscnobfluct}), where we neglect the second term in the numerator (kurtosis), and Eq.~(\ref{v2v3cumul}) to express the skewness as a function of $\Gamma_S$. 
Using Eqs.~(\ref{v32}), (\ref{v22}) and (\ref{v24gauss}), we obtain: 
\begin{equation}
  {\rm nsc}_{2,3}\{4\}= \frac{2\Gamma_S}{\kappa_2\kappa_3} \frac{v_3\{2\}v_2\{4\}^2\sqrt{v_2\{2\}^2- v_2\{4\}^2}}{v_2\{2\}^2}.
   %+\Gamma_K v_2\{2\}^2- v_2\{4\}^2
  \label{nscng}
\end{equation}
One sees that the effect of the intensive skewness is enhanced by a factor  $1/\kappa_2\kappa_3$, which comes from the fact that there are more powers of $\varepsilon_n$ in the denominator than in the numerator of Eq.~(\ref{v2v3cumul}). 
The only quantity that can be inferred from ${\rm nsc}_{2,3}\{4\}$ is the ratio $\Gamma_S/(\kappa_2\kappa_3)$. 
Similarly, for the fluctuations of a single harmonic $v_n$,  the quantities that can be inferred from data are the intensive skewness and kurtosis of $\varepsilon_n$ fluctuations, divided by $\kappa_n^2$. 
In Refs.~\cite{Alqahtani:2024ejg,Roubertie:2025qps},  $\varepsilon_n$ fluctuations were borrowed from a generic parametrization, the elliptic-power distribution~\cite{Yan:2014afa}. 
The response coefficients $\kappa_n$ were extracted from ATLAS data on $v_n$ fluctuations, and found in good agreement with the hydrodynamic values Eq.~(\ref{trajectum}). 
Here, we assume instead that $\kappa_n$ is given by Eq.~(\ref{trajectum}), and we extract $\Gamma_S$ from ATLAS data.  

We fit ATLAS data on ${\rm nsc}_{2,3}\{4\}$ using Eq.~(\ref{nscng}), where we take $v_2\{2\}$, $v_2\{4\}$ and $v_3\{2\}$ from the same analysis (Fig.~\ref{fig:atlasfits} (c) and (d)).  
Note that $v_2\{4\}$ is undefined for the largest values of $N_{ch}$ and $E_T$, which are therefore excluded from the fit. 
The only fit parameter is $\Gamma_S$. 
We carry out a combined fit of the data for $N_{ch}> 750$ (Fig.~\ref{fig:atlasfits} (e)) and $E_T> 950$~GeV (Fig.~\ref{fig:atlasfits} (f)), encompassing roughly the 40\% most central collisions. 
The reason why we exclude peripheral collisions is that the response coefficients may decrease significantly with respect to the values in Eq.~(\ref{trajectum}), due to viscous damping. 
The fit is shown as a dashed line. 
It is of better quality as a function of $E_T$ than as a function of $N_{ch}$, a point to which we come back in Sec.~\ref{s:centrality}. 
The value of the fit parameter is $\Gamma_S\approx -2.1$, in good agreement with the Monte Carlo simulations in Fig.~\ref{fig:trento} (a). 

Eq.~(\ref{nscng}) cannot explain the positive sign of $ {\rm nsc}_{2,3}\{4\}$ in ultracentral collisions. 
This behaviour results from impact parameter fluctuations, which we now take into account. 

\section{Impact parameter fluctuations}
\label{s:centrality}

The centrality resolution is not perfect, and events with the same $N_{ch}$ (or $E_T$) have slightly different impact parameters.\footnote{The effect of centrality fluctuations on ${\rm nsc}_{2,3}\{4\}$ was first studied by Gardim {\it et al.\/}~\cite{Gardim:2016nrr} who pointed out that the result is sensitive to the width of the centrality bin. This implies that the first results from ALICE~\cite{ALICE:2016kpq}, which have wide bins, cannot be compared directly with those of ATLAS~\cite{ATLAS:2019peb} which have much thinner bins.}
Now, the parameters characterizing eccentricity fluctuations, $\bar\varepsilon_2$, $\sigma_{\varepsilon_2}$, $\sigma_{\varepsilon_3}$, depend on the magnitude of impact parameter\footnote{Throughout this paper, we neglect correlations between $\varepsilon_n$ and  $N_{ch}$ (or $E_T$)  at fixed $b$. Their effects were studied in detail in Ref.~\cite{Alqahtani:2024ejg} and found to be small.} $b$ or, equivalently, on the centrality fraction $c\equiv \pi b^2/\sigma_{\rm PbPb}$, where $\sigma_{\rm PbPb}$ is the total inelastic Pb+Pb cross section. 
$c$ is the {\it true\/} centrality, which fluctuates relative to the experimentally-determined centrality~\cite{Das:2017ned}. 

Hence, the averages over events appearing in Eq.~(\ref{defnsceps}) are done in two steps: 
First, one averages over quantum fluctuations at fixed $c$, and the various moments are expressed as a function of $\bar\varepsilon_2$, $\sigma_{\varepsilon_2}$, $\sigma_{\varepsilon_3}$. 
The moments of order $2$ are given by Eqs.~(\ref{eps22}) and (\ref{eps32}), while the moment of order $4$ is obtained using Eqs.~(\ref{4moment}) and (\ref{v2v3cumul}): 
\begin{equation}
\langle \varepsilon_2\varepsilon_2^* \varepsilon_3\varepsilon_3^*\rangle =
(\bar\varepsilon_2^2+\sigma_{\varepsilon_2}^2)\sigma_{\varepsilon_3}^2
+\sigma_{\varepsilon_3}^3\sigma_{\varepsilon_2}\left(2\Gamma_S \bar\varepsilon_2^2+
\Gamma_K\sigma_{\varepsilon_2}^2\right).
\end{equation}
Then, one averages the moments over $c$. 
For Gaussian fluctuations ($\Gamma_S=\Gamma_K=0$), $ {\rm nsc}_{2,3}\{4\}$ no longer vanishes, because the average of the product does not coincide with the product of averages after averaging over $c$. 
Since the magnitudes of $\bar\varepsilon_2$, $\sigma_{\varepsilon_2}$ and $\sigma_{\varepsilon_3}$ all increase with $c$, the average of the product is {\it larger\/} than the product of averages, which explains the positive sign of  $ {\rm nsc}_{2,3}\{4\}$ in ultracentral collisions.  

We now evaluate centrality fluctuations quantitatively. 
The probability distribution of $c$ at fixed $N_{ch}$ (or $E_T$) is obtained by applying Bayes' theorem: $P(c|N_{ch})=P(N_{ch}|c)/P(N_{ch})$~\cite{Das:2017ned}. 
$P(N_{ch}|c)$ is assumed to be Gaussian, and the parameters of the Gaussian are obtained by fitting the histogram of $N_{ch}$ as a superposition of Gaussians (solid lines in Fig.~\ref{fig:atlasfits} (a) and (b)). 
The distributions  $P(c|N_{ch})$ and $P(c|E_T)$ are displayed in Fig.~5 of Ref.~\cite{Roubertie:2025qps}.\footnote{The width of $N_{ch}$ and $E_T$ fluctuations is directly inferred from data only for $c=0$, and one needs to model its variation with  $c$. It can be characterized by a single parameter $\gamma$~\cite{Samanta:2023kfk}, defined as a double ratio (variance/mean) between peripheral and central collisions~\cite{Roubertie:2025qps}. We use the same value  $\gamma=3.5$ as in Ref.~\cite{Roubertie:2025qps}, but our results only depend marginally on $\gamma$, because centrality fluctuations only matter for the largest values of $N_{ch}$ and $E_T$, corresponding to $c\approx 0$.}
The centrality distribution is narrower at fixed $E_T$ than at fixed $N_{ch}$, which means that $E_T$ is a better centrality estimator. 

The next step is to reconstruct the variation of $\bar\varepsilon_2$, $\sigma_{\varepsilon_2}$ and $\sigma_{\varepsilon_3}$ with $c$. 
We carry out this reconstruction following the same method as in Ref.~\cite{Roubertie:2025qps}, by fitting the cumulants of the distribution of $v_2$ and $v_3$ (lines in Figs.~\ref{fig:atlasfits} (c) and (d)).\footnote{Note that this reconstruction includes the leading non-Gaussian corrections to the fluctuations of $\varepsilon_2$ and $\varepsilon_3$. These corrections have a small effect.}  
Eventually, the effect of centrality fluctuations is mostly visible in  ultracentral collisions. 
They explain in particular why the fourth cumulant of the distribution of $v_2$ changes sign, resulting in an undefined $v_2\{4\}$~\cite{Zhou:2018fxx,Alqahtani:2024ejg}, as pointed out in Sec.~\ref{s:nongaussian}. 

We finally average the moments over $c$ in order to evaluate  $ {\rm nsc}_{2,3}\{4\}$. 
We first carry out the calculation by assuming Gaussian flow fluctuations, i.e., with $\Gamma_S=\Gamma_K=0$. 
The result is displayed as a thin full line in Fig.~\ref{fig:atlasfits} (e) and (f). 
One sees that impact parameter fluctuations alone explain the rise and fall of ${\rm nsc}_{2,3}\{4\}$ in ultracentral collisions, as a function of both centrality classifiers. 
They always result in a positive  ${\rm nsc}_{2,3}\{4\}$. 
Its value is larger for fixed $N_{ch}$ than for fixed $E_T$ because centrality fluctuations are broader. 

The negative values of ${\rm nsc}_{2,3}\{4\}$ observed for most centralities can only be explained by taking into account non-Gaussian fluctuations. 
We start by including the skewness. We carry out a one-parameter fit where $\Gamma_S$ is the fit parameter and $\Gamma_K$ is set to zero.\footnote{Note that in practice, the fit parameters are $\Gamma_S/(\kappa_2\kappa_3)$ (and $\Gamma_K/(\kappa_2\kappa_3)$ for the two-parameter fit discussed below), as in Eq.~(\ref{nscng}).} 
It is shown as a dash-dotted line in Fig.~\ref{fig:atlasfits} (e) and (f). 
Agreement with data is now excellent from mid-central all the way up to ultracentral collisions.  
Note that, for peripheral collisions, the quality of the fit is only marginally improved by taking into account centrality fluctuations. 

We finally carry out a two-parameter fit, allowing for a non-zero (but constant) $\Gamma_K$, shown as a solid line. 
The quality of the fit is improved for central collisions which drive the value of $\Gamma_K$. 
The best-fit value is  negative and close to that of $\Gamma_S$. 
However, the improvement brought by this second parameter is marginal, and we do not expect $\Gamma_K$ to be independent of centrality, based on the simulations in Fig.~\ref{fig:trento} (b). 

Therefore, the only meaningful result of our fit is the value of $\Gamma_S$, and we use the differences between the two fits to estimate the error, shown as a band in Fig.~\ref{fig:trento} (a).\footnote{The lower limit of the band is the value returned by the 1-parameter fit, and the upper limit is the value returned by the 2-parameter fit.}
The best-fit value, $\Gamma_S\approx -2$, lies in the middle of our Monte Carlo simulations. 
It is interesting to note that this observable could potentially shed light on the nucleon width $w$, whose preferred value has gone up from 0.5~fm to 1~fm and then back to the original value, as recounted by Giacalone~\cite{Giacalone:2022hnz}. 
Interestingly, the correlation between the transverse momentum and $v_n^2$ was found to display a similar sensitivity to $w$~\cite{Giacalone:2021clp}, with larger $w$ resulting in more negative correlations. 

\section{Summary and perspectives}
\label{s:discussion}

We have shown that the correlation between elliptic flow and triangular flow in nucleus-nucleus collisions is mostly driven by the small non-Gaussianity of initial anisotropy fluctuations, in the form of a ``mixed skewness'' which is the correlation between the reaction plane ellipticity $\varepsilon_{2,x}$ and the square of the triangularity  $|\varepsilon_3|^2$. 
This mixed skewness is described to a good approximation by a single intensive measure $\Gamma_S$ which depends weakly on centrality. 
The value of $\Gamma_S$ inferred from ATLAS data is remarkably close to that predicted by Monte Carlo models of the initial state. 

Impact parameter fluctuations play a crucial role in central collisions~\cite{Alqahtani:2024ejg}: 
They explain quantitatively the positive correlation between $v_2$ and $v_3$, for two different centrality estimators. 
We evaluate them in a data-driven way, by using the measured distribution of the centrality estimator ($P(N_{ch})$ or $P(E_T)$) to infer the magnitude of impact parameter fluctuations.  
Note that state-of-the-art models~\cite{Bernhard:2016tnd,Nijs:2020roc,JETSCAPE:2020mzn} of heavy-ion collisions do not use the distribution of the centrality estimator in the calibration. 
Therefore, there is no guarantee that they have the right  magnitude of impact parameter fluctuations. 

Above 5\% centrality, impact parameter fluctuations play a modest role relative to the non-Gaussianity, thanks to the fine centrality binning implemented by ATLAS. 
Further refining the centrality binning would be of little help: 
The intrinsic centrality resolution is of order 2-3\% at the LHC~\cite{Das:2017ned}.\footnote{This is the full-width at half maximum. The distribution of the true centrality for a fixed value of the centrality estimator is a Gaussian of width 1.2\% for $N_{ch}$ and 0.9\% for $E_T$ in central collisions.}
Therefore, the magnitude of impact parameter fluctuations in a centrality bin is not significantly reduced for bins smaller than 1\%. 

We explain ATLAS data from ultracentral collisions through $\approx 40\%$ centrality, both as a function of $N_{ch}$ and $E_T$, with a single parameter. 
We fail to reproduce data for more peripheral collisions, where the measured anti-correlation keeps increasing. 
A natural explanation of this discrepancy is that the hydrodynamic response $\kappa_n$ decreases for peripheral collisions due to viscous damping. 
We have neglected this effect, which potentially explains a larger anti-correlation according to Eq.~(\ref{nscng}). 
Note that the increase seen by ATLAS is not observed at lower energies~\cite{STAR:2018fpo}  where the anti-correlation decreases for the most peripheral collisions. This difference between STAR and ATLAS may be due to the wider centrality binning~\cite{Gardim:2016nrr}, rather than the different collision energy. 
Schenke {\it et al.\/} point out that agreement with STAR data is improved for peripheral collisions if the hydrodynamic evolution is replaced by a transport calculation in the hadronic phase~\cite{Schenke:2019ruo}. 
This detailed microscopic modeling automatically takes into account the damping of $\kappa_n$ in peripheral collisions.  

Our approach can be extended to different collision systems, in particular O+O collisions~\cite{Brewer:2021kiv} for which data were taken in July 2025 at the LHC collider. 
Based on dimensional arguments, one expects that the magnitude of $\Gamma_S$ should be similar. 
This can be checked in detail through initial-state simulations~\cite{YuanyuanWang:2024sgp,Zhang:2024vkh,Prasad:2024ahm,Loizides:2025ule}. 
We expect that the kurtosis, whose effect is small in Pb+Pb collisions, will play a larger role in O+O collisions, because the system is smaller, and also because the reaction plane anisotropy is not as strong, so that the effect of the skewness is decreased relative to that of the kurtosis. 
In proton-nucleus collisions, the effect of the skewness should be negligible with respect to that of the kurtosis because initial anisotropies are solely due to fluctuations~\cite{Demirci:2021kya,Demirci:2023ejg}. 
Note, finally, that ${\rm nsc}_{2,3}\{4\}$ becomes positive for low-multiplicity collisions in a variety of systems~\cite{CMS:2017kcs,ALICE:2019zfl}, including proton-proton collisions where this reveals information about the proton substructure~\cite{Albacete:2017ajt}. 

Extending our study to symmetric cumulants involving higher harmonics, such as $v_4$ and $v_5$, is not straightforward because they are not determined solely by linear response to initial anisotropies, due to non-linear response terms~\cite{Gardim:2011xv,Teaney:2012ke,Qian:2016fpi,ALICE:2017fcd,Horecny:2025mak}. 
But the methodology, which is to identify generic non-Gaussianities through scaling arguments, should also apply for higher harmonics. 
Finally, our approach can be generalized to higher-order correlations between $v_2$ and $v_3$, which have been thoroughly investigated by the ALICE Collaboration~\cite{ALICE:2021adw,ALICE:2021klf,ALICE:2023lwx}.

\begin{acknowledgments}
M. Alqahtani acknowledges the support of the Research Mobility Program of the French Embassy in Riyadh, which helped to initiate this work. He also acknowledges that this work benefited from State aid under France 2030 (P2I -Graduate School Physique) bearing the reference ANR-11-IDEX-0003.
\end{acknowledgments}

\end{document}